\documentclass[pra, aps, 10pt, twocolumn, floatfix]{revtex4-1} %%

\usepackage {graphicx}
\usepackage {amssymb}
\usepackage {amsmath}

\begin{document}

%%%%%%%%%%%%%%%%%%%%%%%%%%%%%%%%%%%%%%%%%%%%%%%%%%%%%%%%%%%%%%%%%%%%%%%%%%
\title{All-optical control of unipolar pulse generation in resonant medium with nonlinear field coupling}

\author{A.V. Pakhomov$^{1,2}$, R.M. Arkhipov$^{3,4,5}$, I.V. Babushkin$^{6,7}$, M.V. Arkhipov$^{4}$, Yu.A. Tolmachev$^{4}$ and N.N. Rosanov$^{3,8}$}
\affiliation{$^1$ Department of Physics, Samara University, Moskovskoye Shosse 34, Samara 443086, Russia \\
  $^2$ Department of Theoretical Physics, Lebedev Physical Institute, Novo-Sadovaya str. 221, Samara 443011, Russia \\
  $^3$ ITMO University, Kronverkskiy Prospekt 49, St. Petersburg 197101, Russia \\
  $^4$ St. Petersburg State University, Faculty of Physics, Ulyanovskaya str. 3, Petrodvorets, St. Petersburg 198504, Russia \\
  $^5$ Max Planck Institute for the Science of Light, Staudtstraße 2, 91058 Erlangen, Germany \\
  $^6$ Institute of Quantum Optics, Leibniz University Hannover, Welfengarten 1, 30167, Hannover, Germany \\
  $^7$ Max Born Institute, Max-Born-Strasse 2a, 10117, Berlin, Germany \\
  $^8$ Vavilov State Optical Institute, Kadetskaya Liniya v.o. 14/2, St. Petersburg 199053, Russia
}

\date{\today}

\begin{abstract}
We study optical response of a resonant medium possessing nonlinear 
coupling to external field driven by a few-cycle pump pulse 
sequence. We demonstrate the possibility to directly produce
unipolar half-cycle pulses from the medium possessing an arbitrary 
nonlinearity,  by choosing the proper pulse-to-pulse distance 
of the pump pulses in the sequence. We examine the various ways of 
the shaping of the medium response  
using different geometrical configurations of nonlinear oscillators and 
different wavefront shapes for the excitation pulse sequence. Our 
approach defines a general framework to produce unipolar pulses of 
controllable form.
\end{abstract}


\maketitle

\section{\label{sec:section-1}Introduction}

Generation of ultrashort pulses is the field of active research due to emerging opportunities for real-time control of ultrafast processes \cite{Ramasesha, Peng, Kling, Pfeifer}. Since the optical pulses with few-cycle or even subcycle duration became available, the number of their applications increasingly grows continually enhancing our understanding of fundamental phenomena in atoms, molecules and condensed matter. The shortest possible pulse duration opens up new opportunities for the direct measurement of extremely high-speed dynamical processes important for various branches of physics, chemistry, medicine and biology. The development of new techniques for generation and control of the ultrashort pulses is therefore crucial for the advancement of modern optical and material science.

Compared with the few-cycle pulses, half-cycle optical pulses possess an exceptional feature of being unipolar, that is the electric field
does not change its sign throughout the pulse duration. Such a field shape gives rise to the light-matter interactions that are not possible using conventional
light pulses. In particular, this property can be an advantage if one wants to control the ultrafast charge motion in the
pump-probe experiments. Specifically, unipolar pulses can efficiently deliver a kinetic momentum to the charged particles in order to control their motion,
for instance, to ionize the atoms or ions in the medium \cite{Zhang2006, Wetzels, Raman} or to measure the quantum dynamics of electron and ionic wavepackets \cite{Zhang2009, Jones, Bensky, Reinhold}. Unipolar pulses can efficiently accelerate the charge particles and thus be used for producing coherent beams for particle injectors and charge-particle accelerating devices \cite{Rau, Hojo}. Over the last years, unipolar pulses have also been proposed for generation of isolated high-intense attosecond pulses \cite{Orlando, Pan, Feng}. It was recently shown that the coherent light-matter interaction with the use of unipolar pulses enables the control of the resonant medium properties (population inversion and polarization gratings) on a sub-cycle timescale \cite{Arkhipov-gratings, Arkhipov-gratings-2}.
Therefore, finding effective ways of unipolar pulse production and control is needed. It is important to note though, that outlined applications
do not require pulses that are strictly unipolar. Effectively one can consider pulses with the central part of constant sign and long tails on each side being opposite in sign
what allows to avoid the difficulties of dealing with zero-frequency component.

To date, subcycle pulses have been generated in the visible, near-infrared, terahertz, x-ray ranges \cite{Hassan, Manzoni, Bartel, Reimann, Babushkin-tz} and recently in mid-infrared \cite{Zheltikov}. Half-cycle pulses have been experimentally obtained from laser-driven plasma in a solid target \cite{Gao} and from a double foil target irradiated with intense few-cycle laser pulses \cite{Wu}. Theoretically unipolar pulses were predicted when an initially bipolar ultrashort pulse propagates in a nonlinear resonant medium \cite{Kozlov, Kalosha, Song, Song-2} and in Raman-active medium in the self-induced transparency regime \cite{Belenov, Belenov-2} or under the excitation by few-cycle pulses \cite{Arkhipov, Arkhipov-2, Arkhipov-3}. 

Ultrashort laser pulse shaping technologies find ever-widening applications in the fields of coherent control, high-field
laser-matter interactions, nonlinear microscopy and spectroscopy, bio-technology and optical data processing \cite{Weiner, Silberberg,
Goswami, Tearney}. All these applications turn to be highly sensitive to the maximal flexibility in the ultrashort pulse
profiling. Today's research mainstream in the generation of ultrabroadband pulses develops towards the sub-cycle waveform synthesis
 that is the coherent combination of pulses generated from different sources covering separate
spectral regions \cite{Manzoni, Reimann, Wirth, Huang, Chia, Cox, Krauss}. The synthesis of ultrashort pulses calls for careful manipulation
 of the parameters of the individual pulses, namely the spectral phase, carrier phase and the relative
 delay between the sources. Ultrabroad spectrum also requires high-accurate spectral phase
and amplitude control to process the ultrashort pulses, what implies the usage of complicated pulse shaping
 systems to manipulate components extended over large bandwidth. Therefore, the possibilities of the arbitrary optical waveform
 generation appear to be technically challenging and are naturally limited by the shaping device processing characteristics. Possibility of the ultrashort
 and particularly the unipolar pulse shaping directly during the generation process thus
seems to be the extremely inviting prospect. We devise in this paper an alternative approach
for the generation of ultrashort waveforms that we hope to make a contribution towards this direction.

In the work \cite{Arkhipov-4} the optical response of a
one-dimensional string made of two-level oscillators with a
periodically varying density excited at the superluminal velocity by
the incident ultrashort pulse was studied. It was shown that such a
system allows to generate Cherenkov radiation of unusual shape;
namely, additional resonant frequency appeared in the radiation
spectrum. Furthermore, in \cite{Pakhomov} this consideration was
extended to the case when the coupling of the oscillators to the field
is nonlinear; a novel concept of unipolar pulse generation was
proposed when the oscillators are excited by a train of few-cycle
pulses.

In this paper, we generalize this approach to the case of an arbitrary
shape of nonlinear coupling of the oscillators to the external field as well as arbitrarily curved  
wavefronts and different geometries of the oscillators arrangement. Special
consideration is devoted to the role of carrier-envelope phase (CEP)
of the pump pulses in the unipolar pulse generation. We show that CEP value can have a profound impact on the
medium optical response and its proper adjustment can be important to 
obtain the unipolar pulses.

As we show, being combined into the spatially extended arrays, such
oscillators can allow direct generation of unipolar pulses of a
controllable waveform. It is made possible through the excitation
velocity varying upon the propagation along the oscillators array in contrast to the
case of the constant excitation velocity
\cite{Arkhipov, Arkhipov-2, Arkhipov-3}. We study in details several
particular spatial arrangements of the oscillators, for instance
oscillators located on a planar string or on a  
disk. Our findings show that this approach has
great possibilities for the unipolar pulse shaping.

The paper is organized as follows. In Section II we derive
the general condition defining the possibility of unipolar pulse
generation in our system. In Sections
III and IV we discuss the most promising examples of unipolar pulse
shaping. Finally, in Section V we
give a brief overview of our findings and present the concluding remarks.

\section{\label{sec:section-2}General condition of the half-cycle pulse generation}
We consider in the following a model of a resonant optical oscillator excited by a few-cycle pump pulse. We assume the excitation field to be linearly polarized which gives us the following equation for the evolution of the medium polarization $P(t)$: 

\begin{equation}
\ddot{P}+\gamma\dot{P}+\omega_{0}^2P=g[E(t)]E(t),
\label{eq1} 
\end{equation}
where $\omega_{0}$ is the oscillator resonant frequency, $\gamma$ is
the damping rate and the function $g[E(t)]$ describes the coupling
strength of the oscillator to the field. Considering that the field
coupling can be anisotropic, Eq.~\eqref{eq1} should be written for
each component of the polarization vector, but with the linearly
polarized electric field these equations have analogous form to
Eq.~\eqref{eq1}. Hence, without loss of generality we can
restrict ourselves to the scalar case described by
Eq.~\eqref{eq1}.

The coupling field function $g[E(t)]$ can be arbitrary. Physically, it can be implemented by several nonlinearity coupled oscillators with strongly different parameters such as effective mass and resonant frequency. For instance, we may consider just two of such oscillators; the first one is excited by high-frequency external field and induces a slow motion of the other oscillator through the nonlinear bonding. By excluding the slow motion, we arrive to a nonlinear equation like Eq.~\eqref{eq1}. Such nonlinear field coupling can be expected, for instance, for nonlinearity coupled localized plasmonic resonances in the metallic nanostructures \cite{Ginzburg, Berkovitch}, hybrid aggregates of organic supramolecular assemblies and inorganic nanocrystals \cite{Qiao, Savateeva}, coupled semiconductor microcavities \cite{Zhang}, double quantum dot heterostructures \cite{Kouwenhoven, Van der Wiel} or other hybrid optical materials. The natural examples of such field coupling include also Raman-active media \cite{Akhmanov, Nibler}.

We suppose that the pump pulse is short comparing to the natural
period of oscillations: $\omega_{0}\tau_{p}\ll 1$. We also assume the
oscillations decay rate $\gamma \ll \omega_{0}$ and consider the
oscillations on the time interval of the order of the natural period, so that oscillator damping can be neglected. Setting then $u(t)=\dot{P}(t)+i\omega_{0}P(t)$, we obtain the following equation:
\begin{equation}
\dot{u}-i\omega_{0}u=g[E(t)]E(t).
\label{eq2} 
\end{equation}

The assumption of short pulse duration allows to suggest the
oscillator to be affected by instantaneous forcing followed by the
free oscillations. With that Eq.~\eqref{eq2} yields for the induced polarization dynamics right after the pump pulse action for $t>0$:

\begin{equation}
u(t)=e^{i\omega_0t}\Big\{u_0+\int_{-\infty}^{+\infty}g[E(t')]E(t')e^{-i\omega_0t'}dt'\Big\}.
\label{eq3} 
\end{equation}

The integral in the right hand of Eq.~\eqref{eq3} is considered to be taken over the whole pulse duration what is indicated by the infinite integration limits. Taking in mind the electric field being real and splitting the complex exponent under the integral sign
into real and imaginary parts, Eq.~\eqref{eq3} gives:

\begin{gather}
P(t)=P_0\sin(\omega_0t+\phi_0)+\frac{\Pi_1}{\omega_0}\sin(\omega_0t)-\frac{\Pi_2}{\omega_0}\cos(\omega_0t),
\label{eq4} \\
\dot{P}(t)=\omega_0P_0\cos(\omega_0t+\phi_0)+\Pi_1\cos(\omega_0t)+\Pi_2\sin(\omega_0t),
\label{eq5} 
\end{gather}
where $P_0$ and $\phi_0$ correspond to the oscillation amplitude and phase at the moment of the excitation pulse arrival and $\Pi_1, \Pi_2$  are given by:

\begin{eqnarray}
\nonumber 
\Pi_1=\int_{-\infty}^{+\infty}g[E(t')]E(t')\cos(\omega_0t')dt', \\
\Pi_2=\int_{-\infty}^{+\infty}g[E(t')]E(t')\sin(\omega_0t')dt'.
\label{eq6} 
\end{eqnarray}

Oscillator is supposed to be initially at a standstill, i.e. $P_0=0$. Taking this, as we showed in recent papers \cite{Arkhipov, Arkhipov-2, Arkhipov-3}, to emit unipolar pulses, the sine response in Eq.~\eqref{eq4} is required. In this case the oscillator emission induced by one few-cycle pulse can be stopped by another identical pulse at the half-period delay in such a way that the emitted field maintains constant sign. Emitted field then has the form of the half-cycle optical pulse. This result holds since the far-field emitted by the oscillator in between the excitation pulses action at an arbitrary point $\overline{r}'$ is given as:

\begin{equation}
\overline{E}(\overline{r}',t)\thicksim\ddot{\overline{P}}(\overline{r},t-|\overline{r}-\overline{r}'|/c)\thicksim\omega_{0}^2\overline{P}(\overline{r},t-|\overline{r}-\overline{r}'|/c),
\label{eq7} 
\end{equation}
and thus, except for the constant factor, the emitted field is proportional to the polarization itself. For that reason, presence of nonzero cosine term in Eq.~\eqref{eq4} necessitates the non-unipolar emitted field due to cosine function that has the varying sign during the first oscillation half-period.

Given that, from Eqs.~\eqref{eq4}-\eqref{eq5} one gets the necessary criterion for the unipolar half-cycle pulse production by means of the proposed method:

\begin{equation}
\Pi_2=\int_{-\infty}^{+\infty}g[E(t')]E(t')\sin(\omega_0t')dt'=0.
\label{eq8} 
\end{equation}

Analysis of Eq.~\eqref{eq8} reveals some rather significant results. Let us consider the excitation pulse possessing the symmetric envelope with respect to the middle of the pulse (e.g., of the Gaussian shape) and an arbitrary phase shift of the carrier:

\begin{equation}
E(t)=E_0e^{-t^2/\tau_{p}^2}\cos(\Omega t+\vartheta_{CE}),
\label{eq9} 
\end{equation}
where $\Omega$  is the central frequency and $\vartheta_{CE}$ stands for the carrier-envelope phase (CEP). Suppose first $\vartheta_{CE}=\pm \frac{\pi}{2}$ implying the driving electric field is the odd function with respect to the middle of the pulse. In that case Eq.~\eqref{eq8} means that field coupling function $g[E(t)]$ has to be necessarily odd as well. For example, field coupling strength may be proportional to the odd degree of the driving field, like it was shown previously in \cite{Pakhomov} for the simplest linear dependence.

\begin{figure}[htpb]
\includegraphics[width=0.9\linewidth]{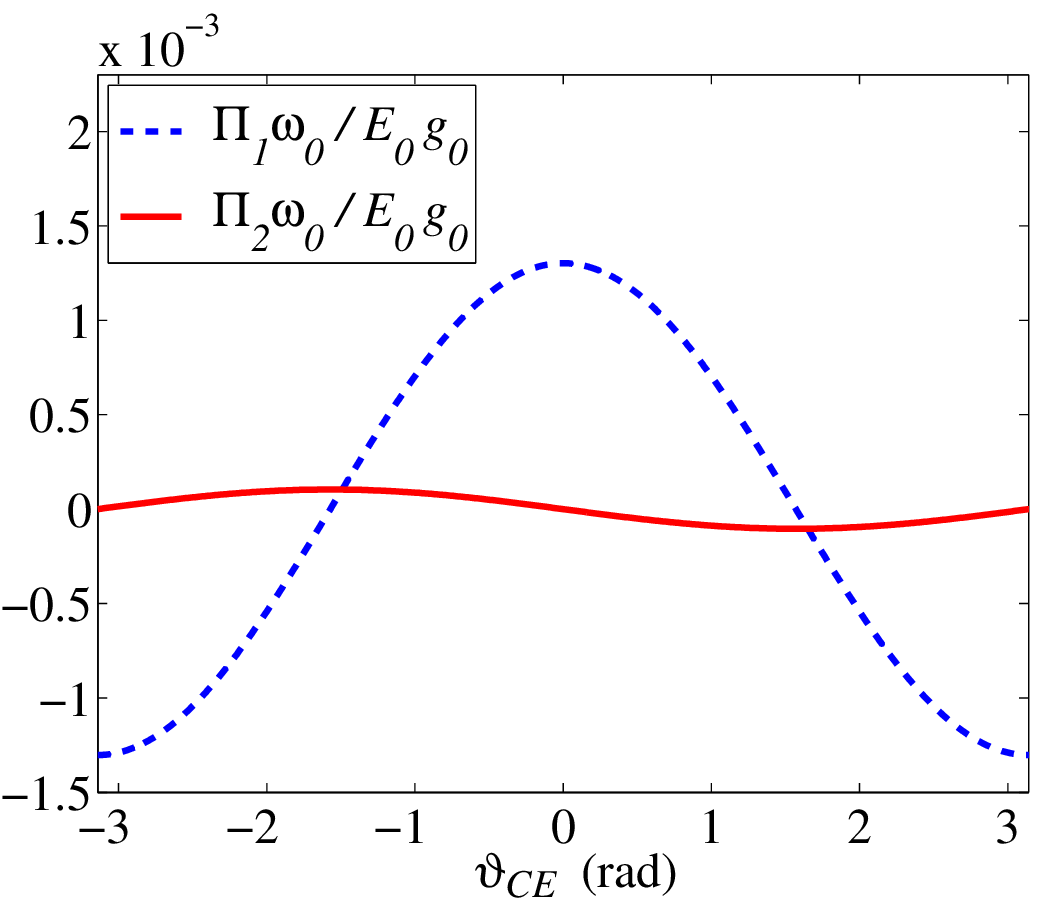}
\caption{The dependence of the linear oscillator response amplitudes $\Pi_1, \Pi_2$ \eqref{pilinear} on the carrier-envelope phase $\vartheta_{CE}$; $\omega_{0}\tau_{p}=0.04$, $\Omega/\omega_0=100$.}
 \label{fig1}
\end{figure}

However, for another carrier-envelope phase $\vartheta_{CE} \ne \pm \frac{\pi}{2}$, the situation turns out different. Assume the most important case $\vartheta_{CE}=0$. As can be seen from Eqs.~\eqref{eq8}, ~\eqref{eq9} the field coupling function $g[E(t)]$  in this case can be of arbitrary power-law dependence on the driving field $E(t)$. This means that even simplest linear oscillator with $g[E(t)]=g_0=const$ can be driven to exhibit sine response and thus to allow the half-cycle pulse generation with accurate adjustment of the carrier-envelope phase. Eqs.~\eqref{eq6} yields the dependence of the linear oscillator response amplitudes $\Pi_1, \Pi_2$ on the carrier-envelope phase $\vartheta_{CE}$:

\begin{eqnarray}
\nonumber 
\Pi_1= \sqrt{\pi} E_0g_0\tau_p e^{-(\Omega^2+\omega_0^2)\tau_p^2/4} \cosh\left(\Omega\omega_0\tau_p^2/2\right) \cos \vartheta_{CE}, \\
\nonumber
\Pi_2=-\sqrt{\pi} E_0g_0\tau_p e^{-(\Omega^2+\omega_0^2)\tau_p^2/4} \sinh\left(\Omega\omega_0\tau_p^2/2\right) \sin \vartheta_{CE}, \\
\label{pilinear}
\end{eqnarray}
that are illistrated in Fig.~\ref{fig1}. Since $\Omega\gg\omega_0$  due to short excitation pulse duration, it is seen from Eq.~\eqref{pilinear} that cosine term amplitude $\Pi_2$ is equal to zero for $\vartheta_{CE}=0$ and $\vartheta_{CE}=\pm \pi$  only, but except for the vicinity of $\vartheta_{CE}=\pm \frac{\pi}{2}$, the sine term amplitude $\Pi_1$ is much greater. Considering that $\omega_{0}\tau_{p}\ll 1$, $\Omega_{0}\tau_{p}\sim 1$, observed relation between response amplitudes $\Pi_1, \Pi_2$ is provided by:

\begin{eqnarray}
\nonumber 
\smash{\displaystyle\max_{\vartheta_{CE}}}  \Pi_2 / \smash{\displaystyle\max_{\vartheta_{CE}}}  \Pi_1=\tanh\left(\Omega\omega_0\tau_p^2/2\right) \approx \Omega\omega_0\tau_p^2/2 \ll 1.
\label{pilin_comp}
\end{eqnarray}

\begin{figure}[htpb]
\includegraphics[width=1\linewidth]{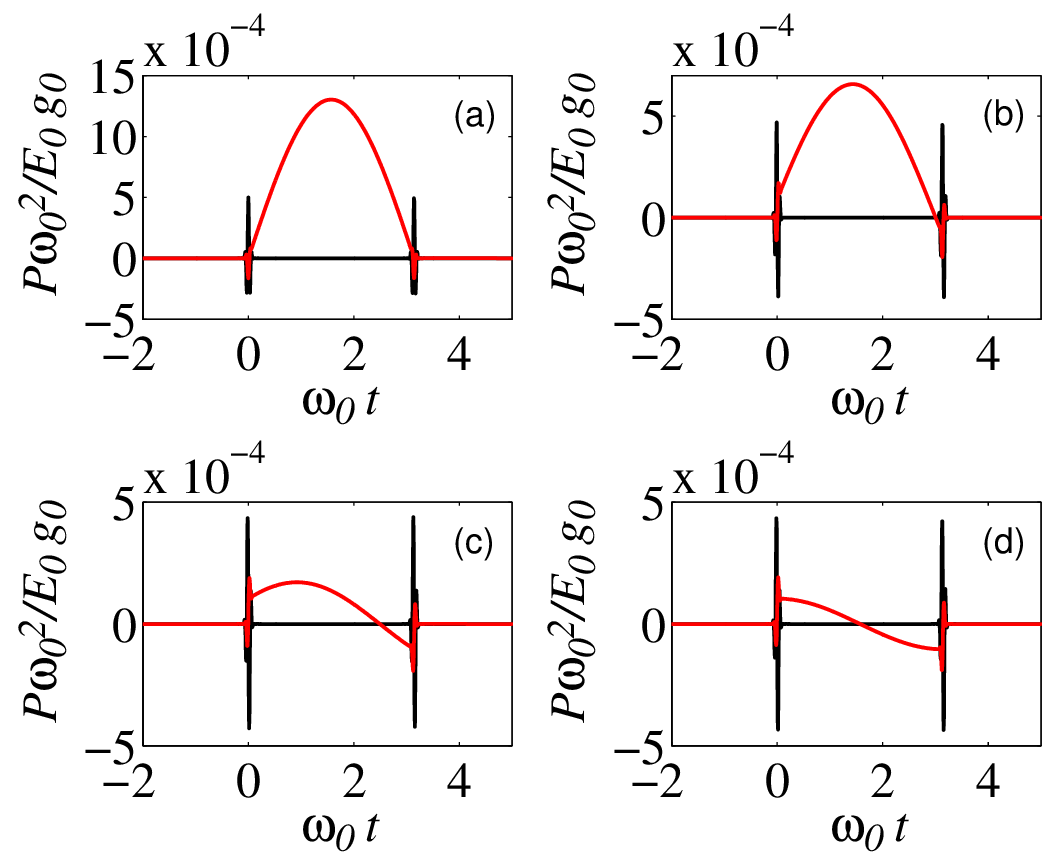}
\caption{The linear oscillator response $P(t)$ (red) and the electric field of the exciting pulses (9) (black) for different values of the carrier-envelope phase: (a) $\vartheta_{CE}=0$; (b) $\vartheta_{CE}=\frac{\pi}{3}$; (c) $\vartheta_{CE}=\frac{7\pi}{15}$; (d) $\vartheta_{CE}=\frac{\pi}{2}$;  $\omega_{0}\tau_{p}=0.04$, $\Omega/\omega_0=100$. The time delay between two excitation few-cycle pulses is equal to $T_0/2$.}
 \label{fig2}
\end{figure}

The resulting oscillator response $P(t)$ is depicted in Fig.~\ref{fig2} (red lines) for different CEP values together with the appropriately scaled electric field of the excitation pulses (black lines). According to Eq.~\eqref{eq7}, the emitted pulse appears to be unipolar when  $\vartheta_{CE}=0$ and $\vartheta_{CE}=\pm \pi$ and contains a certain portion of an opposite sign for other CEP values. Some high-frequency oscillations distorting the emitted pulse shape are inevitably observed during the action of exciting pulses, but we suppose them to be effectively cut off by the appropriate low-pass filter.

%%%%%%%%%%%%%%%%%%%%
%%%%%%%%%%%%

%%%%%%%%%%%%%%%%%%%%
%%%%%%%%%%%%

It should be noted that the oscillation amplitude delivered to the linear oscillator by the excitation pulse is nevertheless substantially smaller than it can be achieved for the oscillator with nonlinear field coupling. The oscillator response amplitudes $\Pi_1, \Pi_2$ for the field coupling function $g[E(t)]=g_1E(t)$ are given as follows:

\begin{eqnarray}
\nonumber 
\Pi_1=\frac{1}{2}\sqrt{\frac{\pi}{2}} E_0^2g_1\tau_p e^{-\omega_0^2\tau_p^2/8} \Big[ 1 + \\
\nonumber 
e^{-\Omega^2\tau_p^2/2} \cosh\left(\Omega\omega_0\tau_p^2/2\right)   \cos 2\vartheta_{CE} \Big],  \\
\nonumber 
\Pi_2=-\frac{1}{2}\sqrt{\frac{\pi}{2}} E_0^2g_1\tau_p e^{-(4\Omega^2+\omega_0^2)\tau_p^2/8} \sinh\left(\Omega\omega_0\tau_p^2/2\right) \cdot   \\
\nonumber 
 \sin 2\vartheta_{CE}. \\  
\label{pinonlin} 
\end{eqnarray}

Eqs.~\eqref{pilinear},~\eqref{pinonlin} yield the following estimate for the maximum values of the response amplitudes:
\begin{eqnarray}
\nonumber 
\frac{\Pi_1^{lin}}{ E_0g_0} : \frac{\Pi_1^{nonlin}}{E_0^2g_1}  \sim  e^{-\Omega^2\tau_p^2/4}.  \\ 
\label{lin-nonlin} 
\end{eqnarray}
Considering that for the pump few-cycle pulse $\Omega_{0}\tau_{p}\sim 1$, one gets from Eq.~\eqref{lin-nonlin} that the nonlinear field coupling enables to significantly increase the 
excitation response. Hence, usage of the linear optical medium seems ineffective although its principal applicability appears to be of crucial importance. 

Another remarkable fact which follows from Eq.~\eqref{pinonlin} is the weak dependence of oscillator response on the CEP, that holds when $e^{-\Omega^2\tau_p^2/2} \ll 1$. As long as this condition is fulfilled, first expression in Eq.~\eqref{pinonlin} converts to:
\begin{eqnarray}
\nonumber 
\Pi_1 \approx \frac{1}{2}\sqrt{\frac{\pi}{2}} E_0^2g_1\tau_p e^{-\omega_0^2\tau_p^2/8}.  \\ 
\label{pinonlin-2} 
\end{eqnarray}

According to Eqs.~\eqref{pinonlin}, \eqref{pinonlin-2} one gets:

\begin{eqnarray}
\nonumber 
\smash{\displaystyle\max_{\vartheta_{CE}}}   \Pi_2 / \smash{\displaystyle\max_{\vartheta_{CE}}}   \Pi_1=\sinh\left(\Omega\omega_0\tau_p^2/2\right) e^{-\Omega^2\tau_p^2/2} \ll 1,
\end{eqnarray}
that is the value of the cosine term amplitude $\Pi_2$ is negligibly small being compared with the sine term amplitude $\Pi_1$, although it is exactly equal to zero just when $\vartheta_{CE}$ is a multiple of $\frac{\pi}{2}$. Thereby, the response of oscillator with nonlinear field coupling is extremely close to sine regardless the CEP value, thus making CEP control insignificant in this case.

For the unipolar pulse shaping purposes, we should consider the summation of the half-cycle pulses from many single oscillators in the certain manner. Thus we arrive to the excitation of the spatially extended arrays composed of specifically arranged optical oscillators. It is interesting to note that being placed into high-Q cavity, such the oscillators arrays serve as the active medium of broad-area passive and active optical systems and can exhibit highly complex spatio-temporal behavior \cite{Babushkin, Loiko, Babushkin-2, Pakhomov-2, Babushkin-3}. When the oscillators array is excited by few-cycle pulses that can generally have different wavefronts form, the emitted unipolar pulse shape is determined by the variation of the excitation velocity over the array. In the simplest case of the linear string excitation at constant velocity \cite{Arkhipov, Arkhipov-2, Arkhipov-4}, the flap-top pulses are produced with only amplitude and duration as the varying parameters. Depending on the array and excitation geometry, we can tune the spatio-temporal profile of the emitted unipolar pulse in the wide limits. In the next sections we examine the possibilities to control the unipolar pulse profile, when we deal with the oscillators with nonlinear field coupling arranged into the spatially extended arrays and excited by the few-cycle pulses having different wavefront forms.

\section{\label{sec:section-3} Planar array of oscillators excited by curved incident waves}
Given the results of the previous section, we consider the possible ways of unipolar pulse generation and control of their spatio-temporal characteristics in the spatially extended arrays of oscillators. 
The optical properties of different nanoemitters and their arrays were actively investigated in recent years. 
Particularly, due to remarkable progress in the development of the metallic nanoparticles fabrication techniques \cite{Kreibig1995, Feldheim2001} 
their ensembles are finding expanding applications as optical waveguides \cite{Maier2003, Quinten1998}, surface enhanced Raman scattering medium \cite{Moskovits1985, Bachelier2004}, 
high quality optical resonators \cite{Alu2006, Citrin2005, Burin}, antennas and detectors \cite{King}. The various types of particles arrangements were considered for these purposes
providing the possibility to adjust the resulting optical characteristics according to the medium geometry.

\begin{figure}[htpb]
\includegraphics[width=1\linewidth]{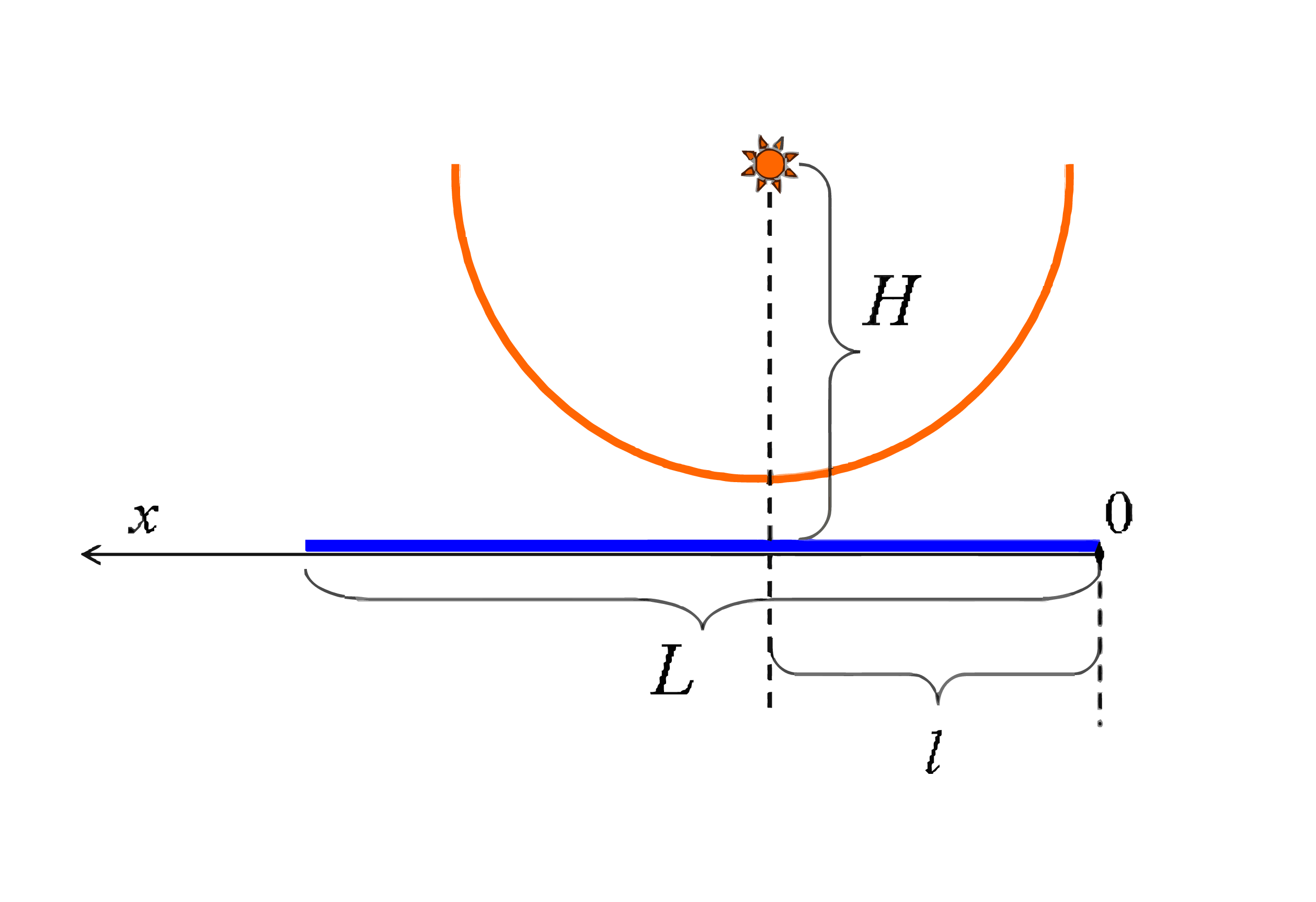}
\caption{A one-dimensional string composed of oscillators with nonlinear field coupling (blue-coloured) is excited by two successive ultrashort light pulses of cylindrical wavefronts (orange-coloured). Oscillators’ emission is observed at a far-distant point on the string midperpendicular.}
 \label{fig4}
\end{figure}

We start with the planar array excited by two few-cycle optical pulses with the $T_0/2$ delay and curved incident wavefronts. As stated above, the oscillator has to exhibit the sine response to the single pump pulse, so the total emission of every oscillator is given by two sine functions with the half-period delay what actually represents half-cycle pulse. Let us first consider the one-dimensional string of oscillators with nonlinear field coupling under the influence of the cylindrical waves (see Fig.~\ref{fig4}). Oscillators’ emission is observed in a point on the string midperpendicular far away from the string at the distance $r \gg L$ or, alternatively, in the focal plane of the focusing lens parallel with the string. For the sake of simplicity, we assume the excitation pulses linearly polarized with the polarization direction orthogonal to the plane of Fig.~\ref{fig4}. Under such assumption, the summation problem is scalar.

%%%%%%%%%%%%%%%%%%%%
%%%%%%%%%%%%%%%%%%%%

\begin{figure}[htpb]
\includegraphics[width=1\linewidth]{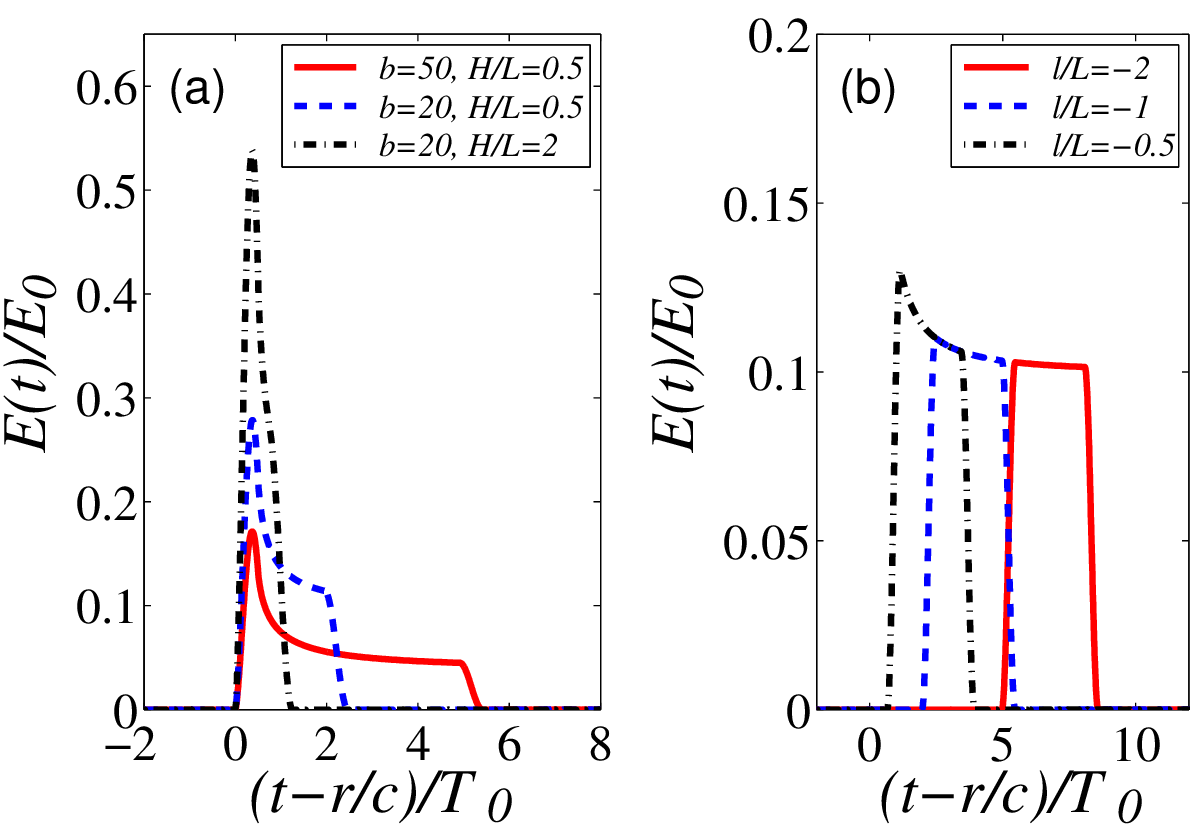}
\caption{Results of the numerical solution of the integral Eq.~\eqref{eq10} for different values of parameters $b$, $\frac{H}{L}$, $\frac{l}{L}$: (a) $l=0$; (b) $b=20, \frac{H}{L}=0.5$.}
 \label{fig5}
\end{figure}

\begin{figure}[htpb]
\includegraphics[width=1\linewidth]{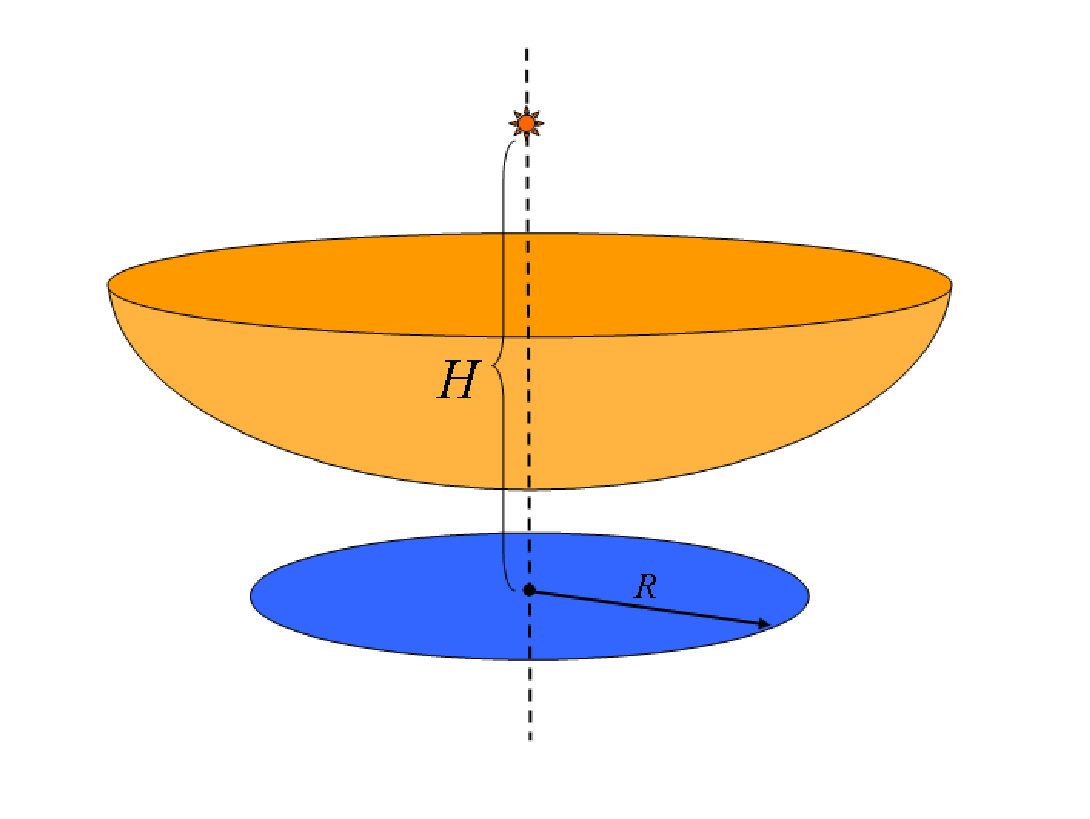}
\caption{Circular disk composed of oscillators (blue-coloured) is excited by two successive few-cycle light pulses with spherical wavefronts (orange-coloured). The medium radiation is observed in a point on the symmetry axis far away from the disk at the distance $r$.}
 \label{fig6}
\end{figure}

\begin{figure}[htpb]
\includegraphics[width=0.9\linewidth]{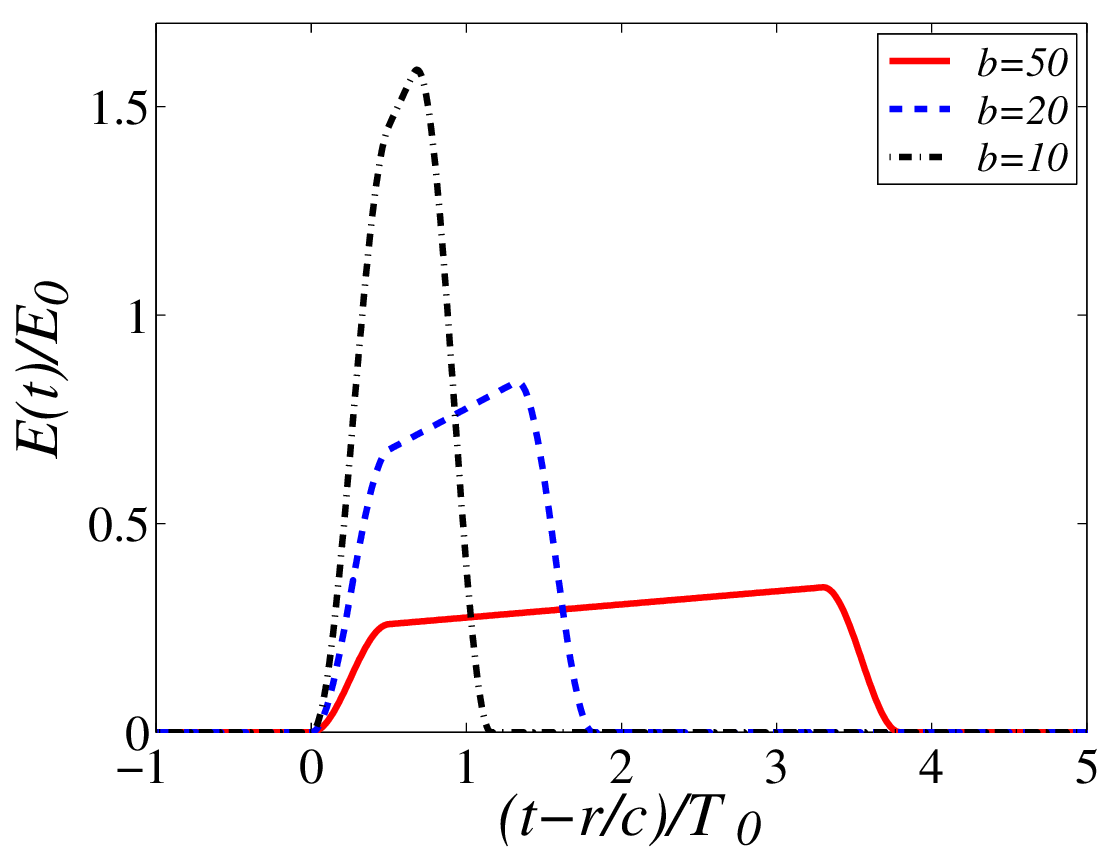}
\caption{Results of the numerical solution of the integral Eq.~\eqref{eq11} for different values of parameter $b$; $\frac{H}{R}=1$.}
 \label{fig7}
\end{figure}

Given Eqs.~\eqref{fig5}-\eqref{fig8} and assuming $\gamma\ll \omega_0$, mathematical expression for the generated pulse shape is written as:

\begin{equation}
E(t)=E_0 \sum_{k=0}^{1} \int_0^L \sin[\omega_0 f_x]\Theta[f_x]dx,
\label{eq10} 
\end{equation}
where $f_x=t-\frac{r}{c}-\frac{\sqrt{H^2+|x-l|^2}-H}{c}-(k-1)T_p$  describes the emission delay from the oscillator located at the point with coordinate $x$, $\Theta$ is the Heaviside step function, $E_0$  is the scaling constant.
The results of the numerical calculation of the integral Eq.~\eqref{eq10} for different values of  parameters $b=\frac{\omega_0 L}{c}$, $\frac{H}{L}$, $\frac{l}{L}$ are plotted in Fig.~\ref{fig5}(a)-(b). The profile of the generated unipolar pulse turns to be monotonically decreasing from its highest level at the leading edge to the lowest at the trailing edge. In such one-dimensional configuration, this shape results from the fact that the intersection point of exciting wavefront moves along the string at the superluminal velocity varying along the string due to wavefront curvature. The final shape asymmetry is determined by this excitation velocity decreasing along the string thus leading to the pulse amplitude decay towards the trailing edge.

%%%%%%%%%%%%%%%%%%%%
%%%%%%%%%%%%%%%%%%%%

Fig.~\ref{fig5}(a) was obtained when the wavefront center is placed right above one of the string ends. In this case excitation wavefronts cross the string at the nearest end at zero angle resulting in infinite instantaneous excitation velocity and the abrupt jump of the pulse amplitude near the leading edge. To make the pulse profile more uniform, one can shift the wavefront center off to the side. Fig.~\ref{fig5}(b) illustrates the unipolar pulses obtained in this way. The pulse shape is now determined by the range of angle values at which the exciting pulses cross the string over its length. When shifting the wavefront center further apart, this angles range decreases leading to the smoothing of the pulse profile. 

To get another pulse shape, we can introduce additional degree of freedom meaning to get varying number of excited oscillators growth rate interfering with the wavefront curvature effects. It will be naturally provided if we turn to two-dimensional geometry and consider the circular disk composed of oscillators and excited by the spherical incident wavefronts (see Fig.~\ref{fig6}). For the sake of simplicity, we assume the excitation pulses linearly polarized with the polarization direction parallel to the circle plane. Array emission is considered being measured in a point on symmetry axis far away from the disk at the distance $r\gg R$ or, alternatively, in the focal plane of the focusing lens parallel with the circle plane.

%%%%%%%%%%%%%%%%%%%%
%%%%%%%%%%%%%%%%%%%%

Mathematical expression for the pulse shape is written in this case as:

\begin{equation}
E(t)=E_0 \sum_{k=0}^{1} \int_0^R \sin[\omega_0 f_{\rho}]\Theta[f_{\rho}] \cdot 2\pi\rho d\rho,
\label{eq11} 
\end{equation}
where $f_{\rho}=t-\frac{r}{c}-\frac{\sqrt{H^2+{\rho}^2}-H}{c}-(k-1)T_p$.

In contrast to the linear array, the observed unipolar pulse Eq.~\eqref{eq11} has now profile monotonically increasing with time (see Fig.~\ref{fig7}; $b=\frac{\omega_0 R}{c}$). Despite the excitation wave moves along the array at gradually decreasing velocity, number of excited oscillators increases faster, thus getting the pulse shape asymmetric towards the trailing edge. 

%%%%%%%%%%%%%%%%%%%%
%%%%%%%%%%%%%%%%%%%%

It should be noted, that taking into account the arbitrary polarization direction of the incident pulses can also provide plenty of alternatives for the pulse shaping. Therefore, polarization degrees of freedom seem to have great potential in this context and deserve special consideration, which we stay outside the scope of current paper.

\section{\label{sec:section-4} Circular array of oscillators excited by plane incident waves}
Another practically relevant way to excite the circular array implies its excitation by the pulses with plane incident wavefronts. We consider once again two few-cycle optical pulses with the $T_0/2$ delay possessing plane wavefronts and propagating at an arbitrary angle $\beta$ to the circle plane. This means the excitation fronts to move in the circle plane at the velocity $V=c/\sin \beta$. Electric field is measured as before at a far distant point at the considerable distance $r$ from the circle: $r\gg R$ or in the focal plane of the focusing lens parallel with the circle plane.

%%%%%%%%%%%%%%%%%%%%
\begin{figure}[htpb]
\includegraphics[width=0.7\linewidth]{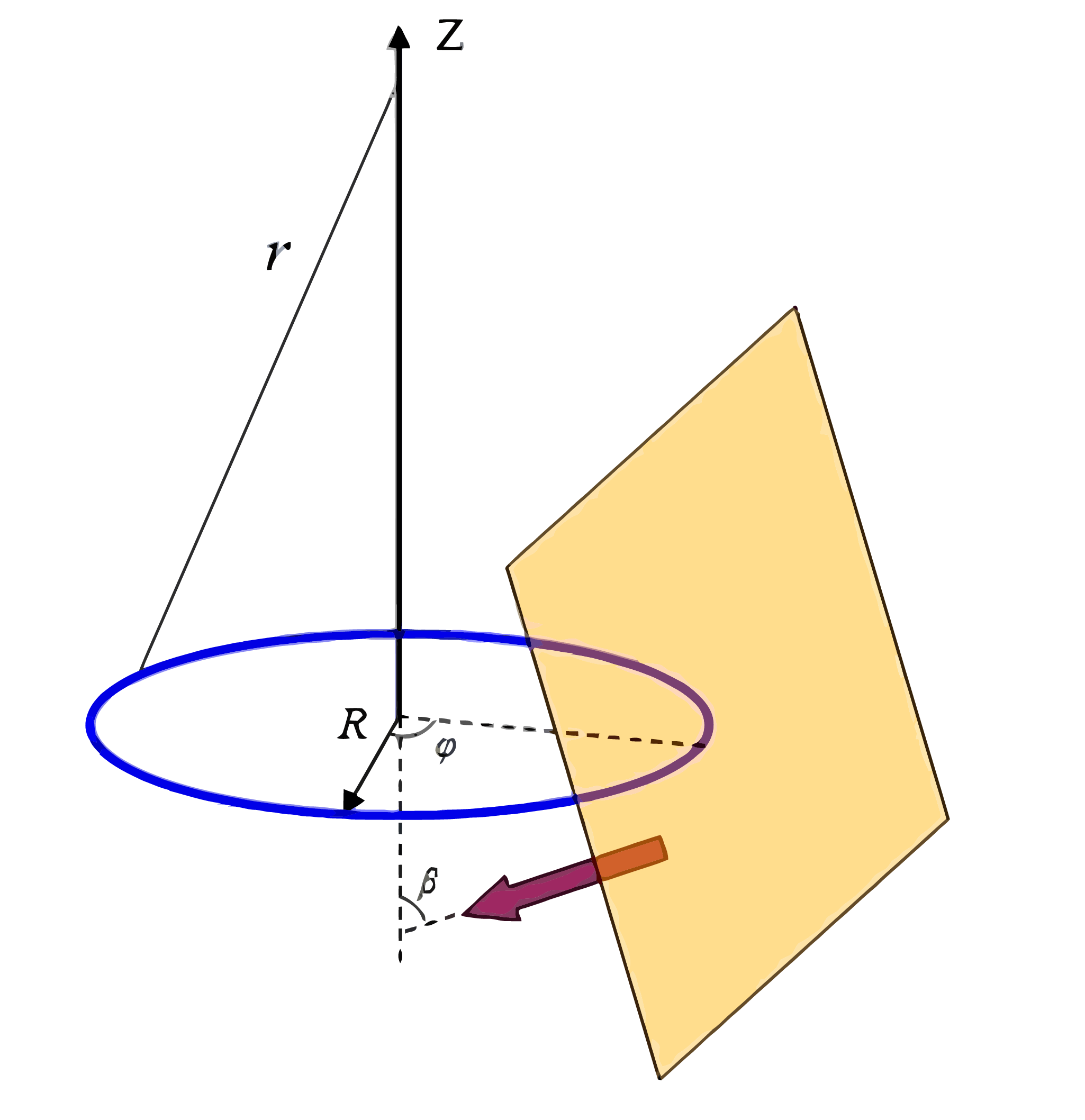}
\caption{Circular string of oscillators (blue circle) is excited by two successive few-cycle light pulses having plane wavefronts and propagating at an angle $\beta$ with the circle plane. Radiation from the medium is observed in a point on $z$ axis far away from the circle at the distance $r$.}
 \label{fig8}
\end{figure}
%%%%%%%%%%%%%%%%%%%%

\begin{figure}[htpb]
\includegraphics[width=1\linewidth]{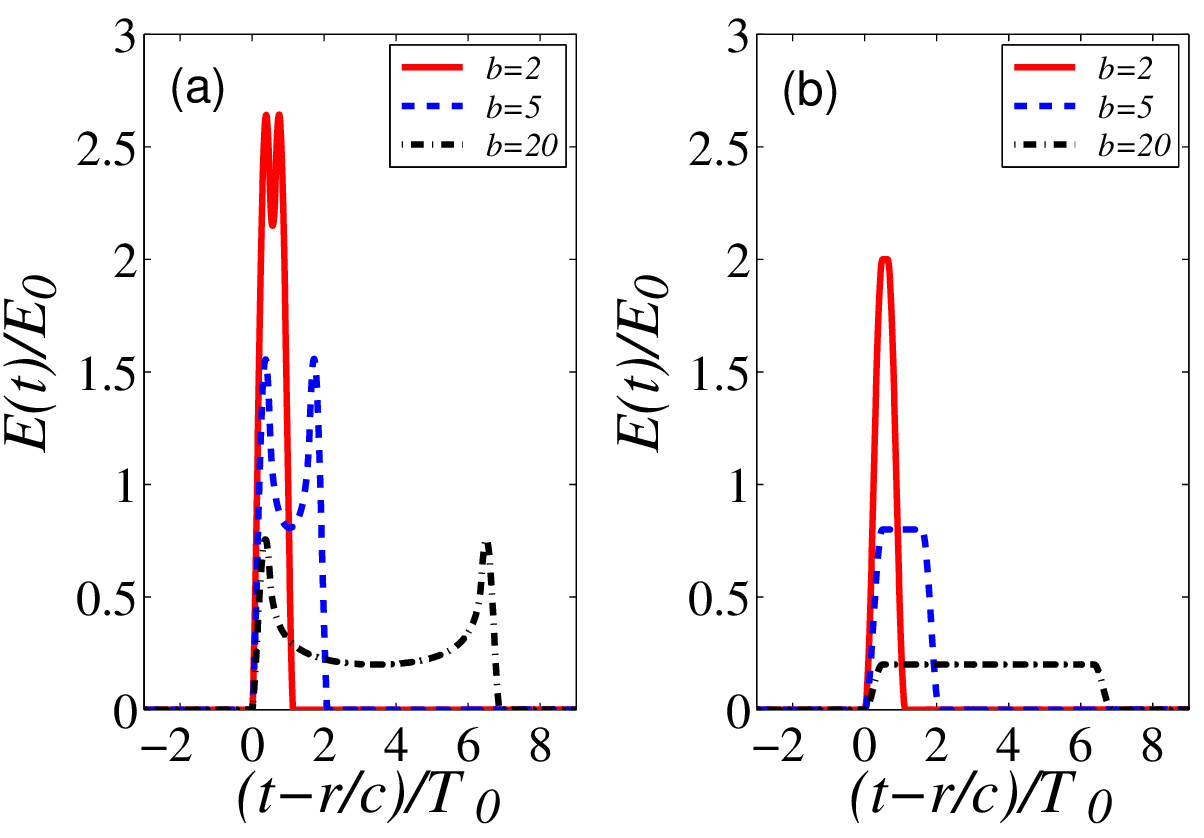}
\caption{Results of the numerical solution of the integral Eq.~\eqref{eq12} for different values of the parameter $b$: (a) with the uniform angular density of the oscillators; (b) with the angular density varying as $N(\varphi)=|\sin \varphi|$.}
 \label{fig9}
\end{figure}

As a first step, we take the oscillators arranged along a circular string of the radius $R$ (see Fig.~\ref{fig8}). The shape of the emitted pulse is expressed now as follows:

\begin{equation}
E(t)=E_0 \sum_{k=0}^{1} \int_0^{2\pi} \sin[\omega_0 f_{\varphi}]\Theta[f_{\varphi}] d\varphi,
\label{eq12} 
\end{equation}
where $f_{\varphi}=t-\frac{r}{c}-\frac{R(1-\cos \varphi)}{V}-(k-1)T_p$  describes the emission delay from the oscillator located at the point with polar angle $\varphi$.
The results of the numerical calculation of the integral Eq.~\eqref{eq12} for different values of the dimensionless parameter $b=\frac{\omega_0 R}{V}=\frac{\omega_0 R}{c}\sin \beta$  are plotted in Fig.~\ref{fig9}(a). Resulting unipolar pulse has symmetric but strongly nonuniform profile with the well-pronounced concave shape. This feature is naturally expected to take place since the media geometry configuration is nonlinear and the intersection point thus moves along the string at varying velocity $V/|\sin \varphi|$. Indeed, the intersection point velocity has the maximum values at $\varphi=0$ and $\varphi=\pi$, resulting in the field jumps near the leading and trailing edges of the pulse. That is why generation of a rectangular pulse does not occur in this case. With decreasing the angle $\beta$ and thus the $b$ parameter, the overall duration of the generated unipolar pulse properly decreases as it takes less time for the excitation pulses to get through whole the circle; the pulse amplitude at the same time correspondingly grows. In the limiting case when the incident wavefronts are parallel to the circle plane $b\to0$, all the particles radiate in-phase thus producing a single half-cycle pulse multiplied by the number of oscillators. 

%%%%%%%%%%%%%%%%%%%%
%%%%%%%%%%%%%%%%%%%%

To smoothen the pulse shape, we may take the particles density varying along the circle in a certain manner. With $N(\varphi)$ the angular density of oscillators’ distribution, Eq.~\eqref{eq12} generalizes to become:

\begin{equation}
E(t)=E_0 \sum_{k=0}^{1} \int_0^{2\pi} \sin[\omega_0 f_{\varphi}]\Theta[f_{\varphi}]N(\varphi) d\varphi.
\label{eq13} 
\end{equation}

Fig.~\ref{fig9}(b) shows an important example of such distribution \eqref{eq13} for $N(\varphi)=|\sin \varphi|$  with producing rectangular-shaped unipolar pulses. Its duration increases with the increase of $b$ while amplitude correspondingly decreases thus keeping the whole pulse area constant.

Finally, we examine the circular disk-shaped array of oscillators analogous to the geometry from the previous section. Equation for the pulse shape is expressed as:

\begin{equation}
E(t)=E_0 \sum_{k=0}^{1} \int_0^{2R} \sin[\omega_0 f_{x}]\Theta[f_{x}] \cdot 2\sqrt{1-{\left|\frac{x}{R}-1\right|}^2} d\left(\frac{x}{R}\right),
\label{eq14} 
\end{equation}
where $f_{x}=t-\frac{r}{c}-\frac{x}{V}-(k-1)T_p$  and $x$ axis goes along the diameter through the disk center.

%%%%%%%%%%%%%%%%%%%%
\begin{figure}[htpb]
\includegraphics[width=0.9\linewidth]{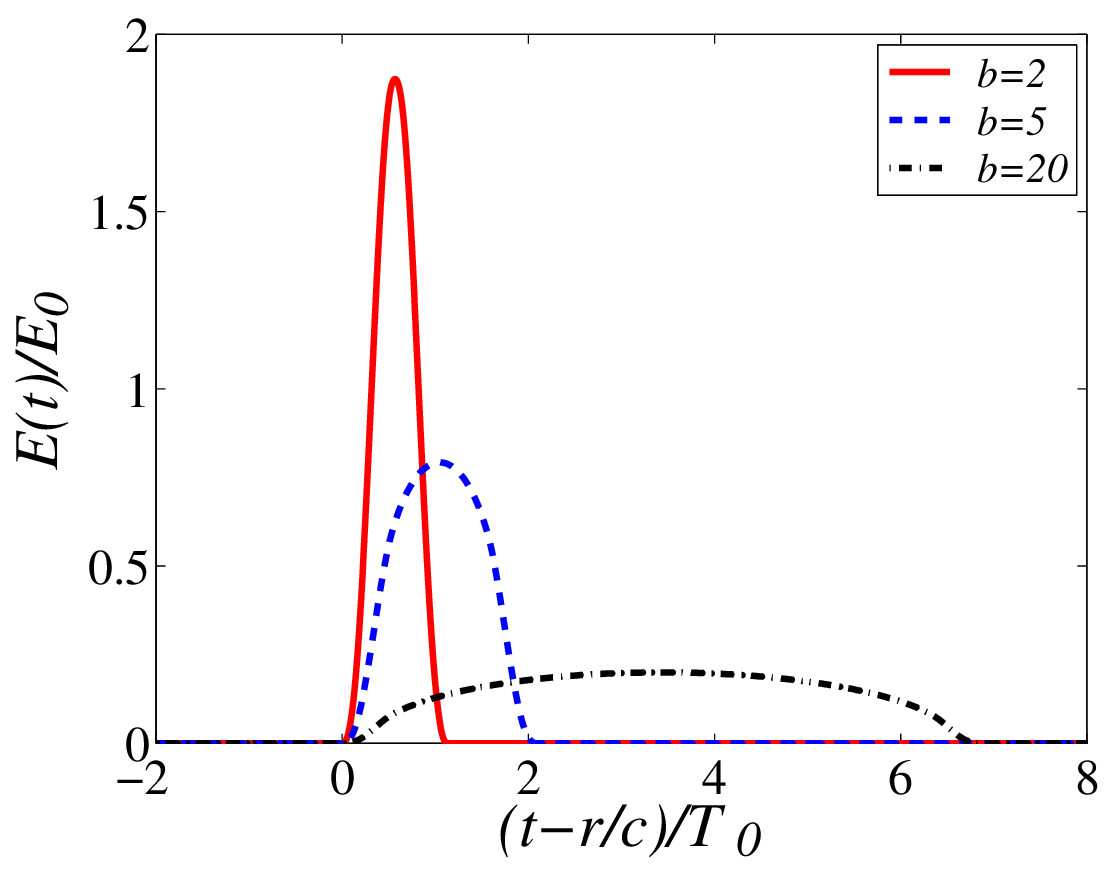}
\caption{Results of the numerical solution of the integral Eq.~\eqref{eq14} for different values of the parameter $b$.}
 \label{fig10}
\end{figure}
%%%%%%%%%%%%%%%%%%%%

Fig.~\ref{fig10} shows the results of the numerical calculation of the integral Eq.~\eqref{eq14} for different values of the dimensionless parameter $b=\frac{\omega_0 R}{V}=\frac{\omega_0 R}{c}\sin \beta$. Since the number of excited oscillators growth rate reaches its peak value over the passing through the disk, resulting unipolar pulse has convex profile in this case. Its duration and amplitude may be inversely varied according to the parameter $b$.

\section{\label{sec:section-5} Conclusion}
We have studied theoretically the optical response of a resonant
medium with the essentially nonlinear field coupling which is excited by
few-cycle pump pulses. The medium was shown to exhibit specific
 response depending on the field coupling
nonlinearity, oscillator geometry and the CEP of the excitation
pulse. Relying on the response specifics, we elaborated the concept of
the half-cycle pulse generation when the oscillator is influenced by a
pair of pulses with the proper time delay, so that the first pulse
initiates the half-cycle pulse response and the second one stops it at
the certain moment. The general criterion was derived allowing to
determine the applicability of this method for the arbitrary
nonlinearity of the field coupling. 

The proposed method allows to produce ultrashort unipolar pulses directly
from the resonant medium without usage of any complicated technique for the
coherent waveform synthesis. Since the only restriction we imposed on the medium
resonant frequency $\omega_{0}$ is its corresponding period being much larger than
the excitation pulse duration, our approach can be applicable for the controllable
generation of half-cycle pulses in the wide frequency range. 
Therefore, the method can be most easily used in terahertz and 
mid-infrared ranges. However, potential extension of proposed
approach for the femtosecond and subfemtosecond range seems not impossible.

To study the possibility of the pulse shaping in our scheme, we
considered the emission from spatially extended arrays of different
geometry as well as different wavefront forms of the excitation
pulses. We have studied in detail several particular cases, namely
when a planar array of oscillators is excited by an incident wave with
a curved front, and a case of a circular array excited by a plane
incident wave. It was shown that modification of the oscillator
density and excitation wavefront allows generation of unipolar pulses
with high variability of pulse shapes: asymmetric ones with
the maxima at the leading or trailing edges and with monotonous
increase or decrease towards the opposite edge, or symmetric ones with either
concave, or convex, or rectangular shape.

\begin{acknowledgments}
R.A. thanks Government of Russian Federation (074-U01); Russian Foundation for Basic Research (16-02-00762); I.B. thanks German Research Foundation (DFG) (project BA 4156/4-1); Nieders. Vorab (project ZN3061).
\end{acknowledgments}

%\bibstyle{apsrev4-1.bst}
\bibliography{optics}

\end{document}